\documentclass{article}

\usepackage[final]{nips_2017}
\usepackage[utf8]{inputenc} 
\usepackage[T1]{fontenc}    
\usepackage{hyperref}       
\usepackage{url} 
\usepackage{booktabs}       
\usepackage{amsfonts}       
\usepackage{nicefrac}       
\usepackage{microtype}      
\usepackage{graphicx}
\usepackage{amsmath}
\usepackage{csquotes}
\usepackage{enumitem}
\usepackage{sidecap}
\usepackage{wrapfig}
\usepackage{caption}
\usepackage{subcaption}
\title{Soc2Seq: Social Embedding meets Conversation Model}

\author{
Parminder Bhatia\thanks{ Work was done when author was working at YikYak Inc. } \\
Amazon Inc.\\
\texttt{parmib@amazon.com} \\
\And
Marsal Gavald\`a \\
Square Inc. \\
mgavalda@gmail.com \\
\And
Arash Einolghozati \\
Amazon Inc. \\
einolghozati@gmail.com \\
}

\begin{document}

\maketitle

\begin{abstract}
 While liking or upvoting a post on a mobile app is easy to do, replying with a written note is much more difficult, due to both the cognitive load of coming up with a meaningful response as well as the mechanics of entering the text. Here we present a novel textual reply generation model that goes beyond the current auto-reply and predictive text entry models by taking into account the content preferences of the user, the idiosyncrasies of their conversational style, and even the structure of their social graph. Specifically, we have developed two types of models for personalized user interactions: a \textbf{content-based conversation model}, which makes use of location together with user information,  and a \textbf{social-graph-based conversation model}, which combines content-based conversation models with social graphs.
\end{abstract}
\section{Introduction}
Yik Yak is a location-based social app, where people can view text and images posted within a five-mile radius. Users talk about a variety of topics on our platform such as sports, politics, entertainment, food, etc. They voice their opinions on these topics in the form of yaks (posts), comments (replies to yaks), and upvotes and downvotes to both yaks and comments.

This user activity results in a flood of events that we then analyze for a variety of purposes, such as categorizing content, building user profiles, and deriving a social graph, to ultimately provide a more personalized and engaging user experience. 



\begin{figure*}[t!]
    \centering
    \begin{subfigure}[t]{0.5\textwidth}
        \centering
        \includegraphics[scale=.15]{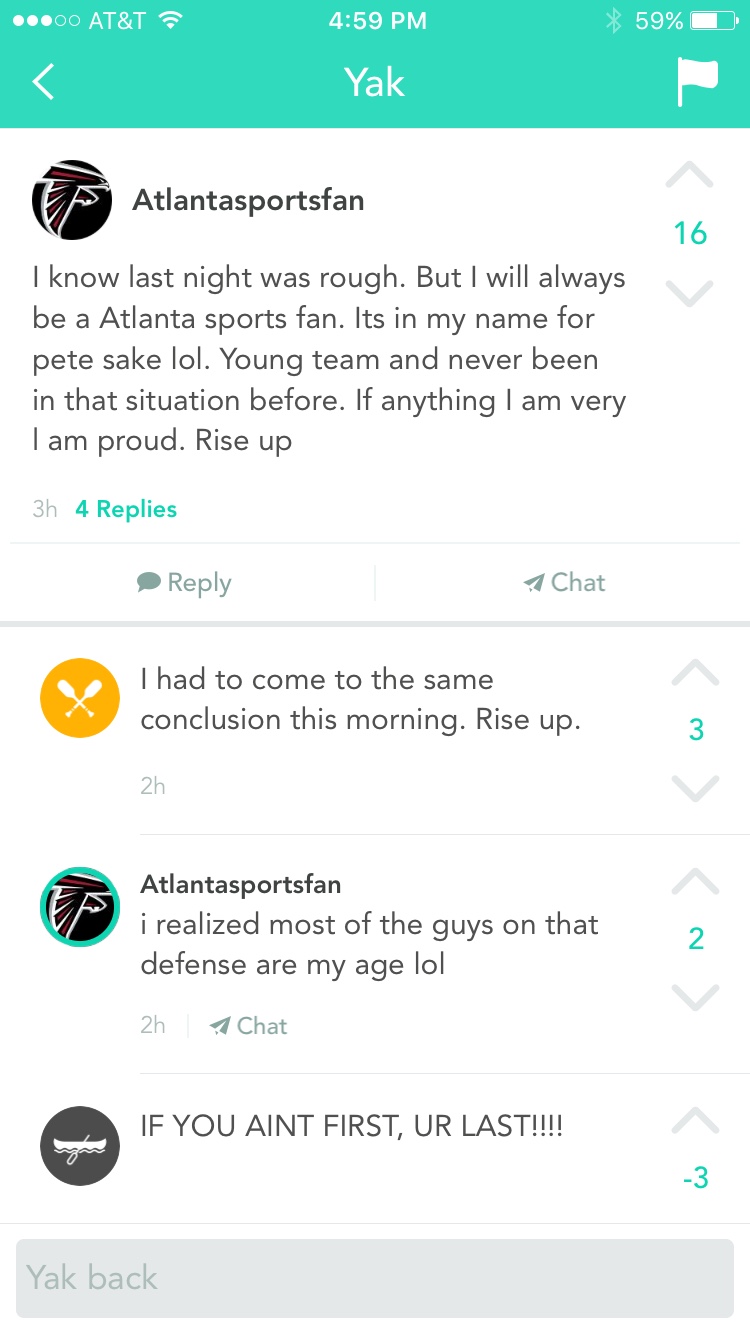}
        \caption{The Yik Yak app displaying a yak (post) along with replies and votes.}
    \end{subfigure}%
    ~ 
    \begin{subfigure}[t]{0.5\textwidth}
        \centering
        \includegraphics[scale=.25]{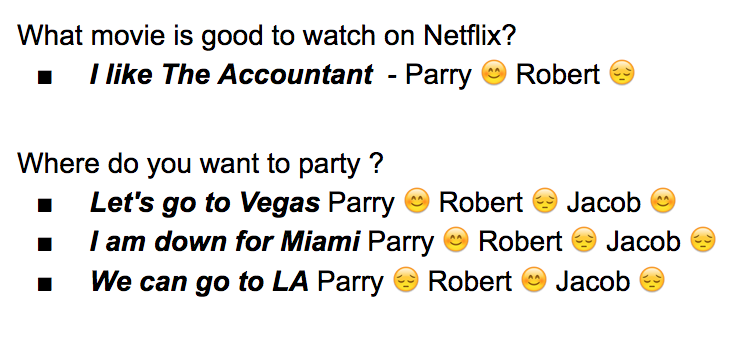}
        \caption{ Examples of why a single suggested reply is not valid for an entire community.}
    \end{subfigure}
    
\end{figure*}

As people discuss a variety of topics on our platform, many different opinions are voiced in the form of upvotes, downvotes, and replies (see Fig. 1). However, typing a reply on a phone is not the most convenient experience, so we have started experimenting with the ability to provide highly personalized reply suggestions, with the hope that it will improve user experience and increase engagement on a per-yak basis as well as on the app overall.

While there has been prior work on using deep learning to generate automated replies, these suggested responses tend to be overly generic, which gave us the impulse to build a novel and robust model that takes into account the location as well as user preferences in the generation of replies.


Let's say, as shown in Fig. 1, the question \textit{What movie is good to watch on Netflix?} is posted. Suggesting just one answer for all responders is not the best way to provide reply options. Thus, we need to build intelligent agents that can be personalized per user or community to give more relevant and accurate responses. The notion of persona or personality can change even for a single user over time or within the same day. For example, a college student may discuss class-related topics on weekday mornings but switch to talk about what events to attend over the weekend.

We present two models for personalization: first, a conversation model consisting of location- and user-based information, and then a social graph-based conversation model that combines the conversation models with social graph information.
\begin{figure*}[!t]
  \includegraphics[scale=.09]{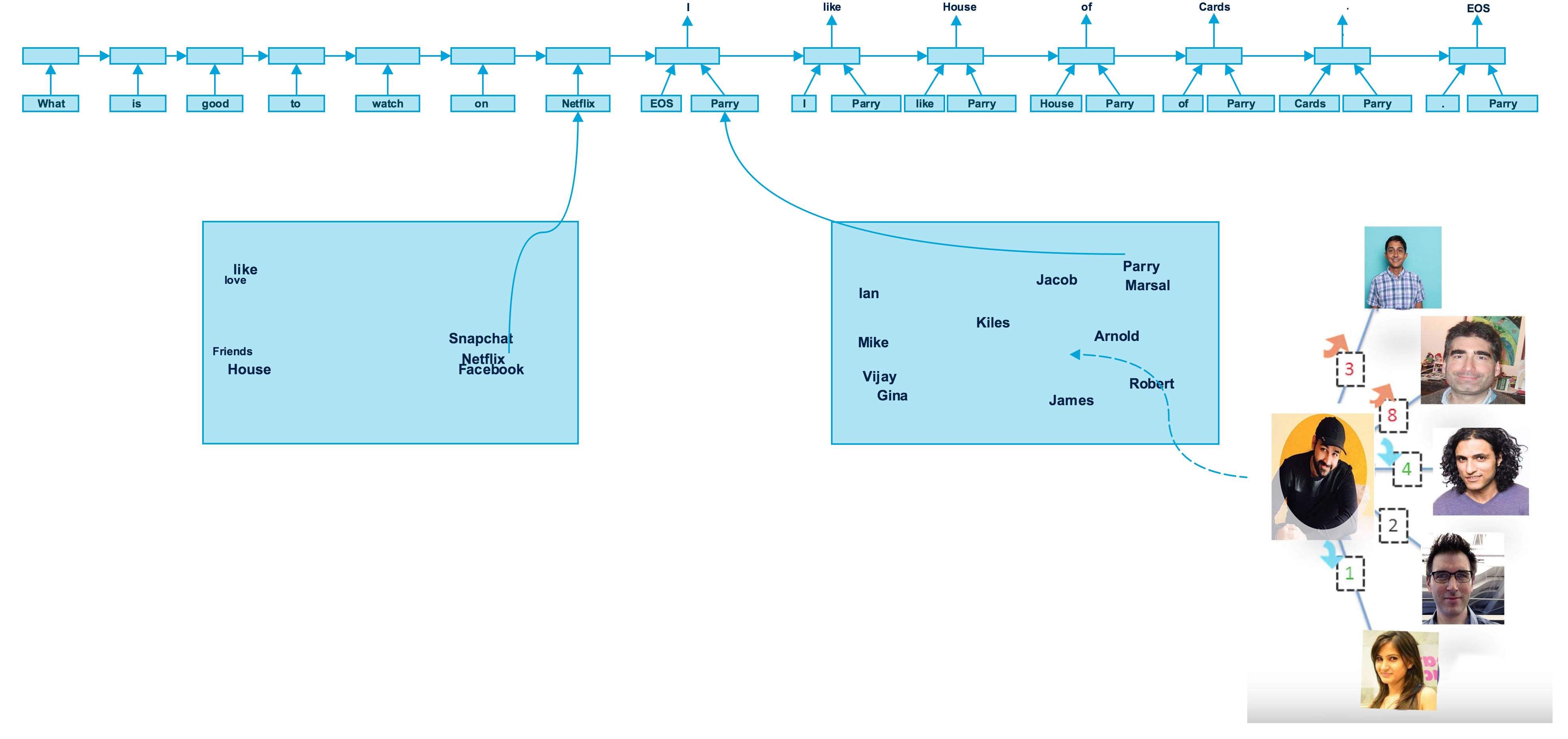}
  \caption{soc2seq: Social Conversation Model.}
  \label{fig:RSUencountered}
\end{figure*}

Main contribution of this paper are
\begin{enumerate}[noitemsep,topsep=0pt]

\item  This represents, to our knowledge, the first time that a neural conversation model has been combined with a social graph to build a more intelligent and personalized agent.
\item Enhanced standard seq2seq LSTM's(Fig. 2) and achieved 15.81\% improvement in ROUGE scores over strong seq2seq baselines.
\item Data used in this model which combines conversation models with location and social attributes.
\item Large scale network used in production for real-world applications such as highly-customized and personalized auto-reply suggestions with results added in appendix section.
\end{enumerate}

\section{Design}
This system consists of Conversation agent, Personalization and Social graphs which interact with one another in a unique way.  Next we talk about each of these components as well as design choices made for them.
\subsection{ Conversation Agent}
\subsubsection{  Conversation Model}
There has also been various work on conversation models or chatbots using neural generative models such as seq2seq. We briefly describe some of the related work in this section.
Chatbots, also called conversational agents or dialog systems, have been studied by a variety of researchers from both academia and industry. We briefly describe the historical work. \\
There are two main classes of conversational models: retrieval-based and generative models. While \textbf{retrieval based models} use a repository of predefined responses and some  heuristic to choose an appropriate response based on the input and the context, \textbf{generative models} go beyond predefined responses and are able to generate new responses from scratch. Here, we discuss SEQ2SEQ \cite{li:persona}, one of the most promising generative models, which is the baseline for our work.

Smart reply \cite{kanan:reply} which is used in Google's auto reply system for emails is one of the practical applications of SEQ2SEQ. Similar to other seq2seq models, the Smart reply system is built on a pair of recurrent neural networks, one used to encode the incoming email and one to predict possible responses. The encoding network ingests the words of the incoming email one at a time and produces a vector. This vector, which Geoff Hinton calls a \textbf{thought vector} \cite{hinton:thought}, captures the gist of what is being said while abstracting from the specific words used.
While these models are fairly easy to train, they have a tendency to produce generic answers for any type of topic or interaction as presented in \cite{li:persona}. Also, they suffer from the same issues as general recurrent models, namely a vanishing gradient when the length of the sentence is too long. An area of research has been \textbf{attention}-based models \cite{badanu:attention}, which emulate how humans give  more importance to certain words in a sentence. Attention mechanism predicts the output using a weighted-average context vector and not just the last state. As an example, in the sentence: \textit{ What is good to watch on TV?} we give more emphasis to \textit{watch} and \textit{TV} as opposed to the other words. For our baseline model, we trained an attention-based SEQ2SEQ model using Yik Yak post and reply data. 
 \subsection{Personalization}
 There has been recent work that incorporates topic \cite{xing:topic} as well as context \cite{ghosh:nlp} in the SEQ2SEQ models in order to generate topic-based responses. While these models do come up with responses that might belong to one domain, they raise the question of whether it is sufficient for an intelligent agent to generate a single answer per topic or context. 
 Li et al. \cite{li:persona} introduced a personalized conversation model to address these issues that broadly consists of two models: the Speaker Model that integrates a speaker-level vector representation into the target part of the SEQ2SEQ model, and a Speaker-Addressee model that encodes the interaction patterns of two interlocutors by constructing an interaction representation from their individual embeddings and incorporates it into the SEQ2SEQ model. These persona vectors are trained on human-to-human conversation data and are used at test time to generate personalized responses. 
 \subsubsection{Personalized Conversation Models}
We introduce two types of conversation models: the \textbf{location based model}, which captures the geo-community aspect of YIKYAK, and the \textbf{user-based model}, which is personalized for each user.
\subsubsection{Location-based model}
The coupling of text with demographic information has enabled computational modeling of linguistic variation, including uncovering words and topics that are characteristic of geographical regions \cite{jacob:geo}. In this paper, we take a step further and introduce a method that extends vector-space lexical semantic models to learn representations of geographically situated language.
Given the location-based nature of Yik Yak, it is  important to incorporate location information in the model. Based on various independent studies, we have found out that the communities can vary significantly from each other in terms of both social connectivity and language usage, whereas the communities are internally homogeneous. In bringing in extra-linguistic information
to learn word representations, our work falls into the general domain of multimodal learning \cite{sri:multi}
that incorporate multiple active modalities (such
as gesture) from a user \cite{sharon:multi}, our primary input is textual data, supplemented with the metadata about the author and the time of authorship. Thus, our first model is a location-based conversation model.

For this approach we developed two persona-based models: 
the \textbf{decoder model}, which captures the personality of the respondent, and the 
\textbf{encoder-decoder model}, which captures the way the respondents adapt their speech to a given addressee. Specifically, we use location embedding for both the encoder and decoder. Details of these models will be discussed in the following section.

Given that each user event in our app is tagged with a latitude and longitude, we have a very robust way of detecting location. Here, we encapsulate the location information using three levels of granularity: county, city, and country. We concatenate the corresponding representations for each level to get the final local embedding. The intuition behind this strategy is that if the data corresponding to a higher level of granularity is sparse, the lower level will provide a stronger signal. 

As an example, the final local embedding for Queens county in New York is given as follows:
\begin{equation}
	\vec{locFinal}_{\,Queens}= [\vec{loc}_{\,Queens} , \vec{loc}_{\,NY} , \vec{loc}_{\,US}]
\end{equation}

We used a final location embedding of size 300 in the model. As in the standard SEQ2SEQ models, we first encode the message $S$ into a vector representation $h_S$ using the source LSTM. Then, for each step in the target side, hidden units are obtained by combining the representation produced by the target LSTM at the previous time step, the word representations at the current time step, and the location embedding.
\begin{equation}
 	\begin{split}
 		& \mathbf i_t = \sigma(\mathbf W^u * \mathbf h_{t-1} + \mathbf I^u * \mathbf [x_t ,\vec{locFinal}_{\,Queens]} ) \\
 		& \mathbf f_t = \sigma(\mathbf W^f * \mathbf h_{t-1} + \mathbf I^f * \mathbf x_t) \\
 		& \mathbf o_t = \sigma(\mathbf W^o * \mathbf h_{t-1} + \mathbf I^o * \mathbf x_t) \\
 		& \mathbf c_t = \tanh(\mathbf W^c * \mathbf h_{t-1} + \mathbf I^c * \mathbf x_t) \\
		& \mathbf m_t = \mathbf f_t \odot \mathbf +  \mathbf i_t \odot \mathbf c_t \\
 		& \mathbf h_t = \tanh(\mathbf o_t \odot \mathbf m_{t-1}) 
 	\end{split}
 \end{equation}
 In our final model, we used an attention-based model which outperforms the standard LSTM. The LSTM defines a distribution over the outputs and sequentially predicts tokens
using a $softmax$ function. Since we want to predict the next word in a sentence, it would be a vector of probabilities across our vocabulary.
Finally, we minimize the average negative log likelihood of the target words:
 \begin{equation}
 	\begin{split}
 	& loss = -\sum_{i=1}^{N}\ln p_{target}
 	\end{split}
 \end{equation}
\subsubsection{User-Based Model}
User-based models are similar to location-based ones with the difference that instead of using the location embeddings, we learn and use the user embeddings based on the conversational interaction between users. Compared with 10,000 location embeddings, we have about 100,000 user embeddings for the model to learn from.
\begin{SCfigure}
  \centering
  \caption{ A toy example for an information network from LINE \cite{tang:line}. Edges can
be undirected, directed, and/or weighted. Vertices 6 and 7 should be placed closely in the low-dimensional space as they are connected through a strong tie. Vertices 5 and 6 should also be placed closely as they share similar neighbors. }
  \includegraphics[scale=.25]{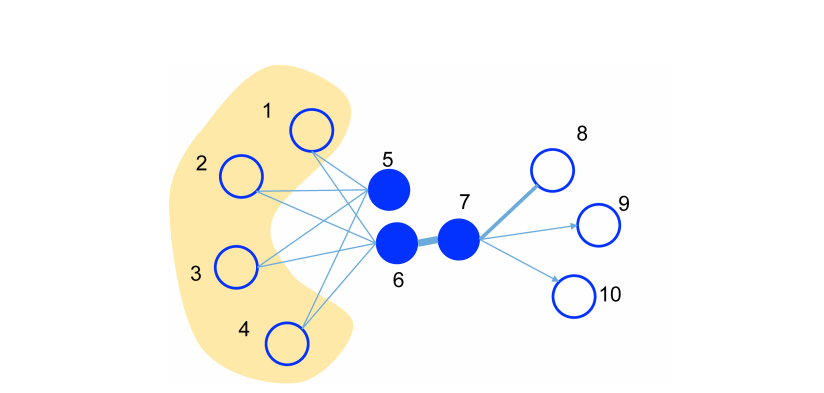}
\end{SCfigure}

\subsection{Social Graph based networks}
Social graphs have numerous applications and are considered the most important ranking factor for functionality such as people discovery (e.g., \enquote{people you may know}) and algorithmic content feed creation (e.g., \enquote{feed personalization}). As we do not have explicit social links such as friends (Facebook) or followers (Twitter) between users, we need to build the latent social graph based on the user interactions within our platform. The idea behind building this model comes from the fact that over time, people will develop a preference for certain users in their community.

Various methods of graph embedding have been proposed in the machine learning literature (e.g., \cite{belkin:eigenmaps,
tenen:dr, cox:md}). They generally perform well on smaller networks. The problem becomes much more challenging in a real-world information network with millions of nodes and billions of edges. In such networks, we seek to efficiently find low-dimensional embeddings that capture the network structure. Fig.  3 shows an illustrative example. Given that the weight of the edge between vertices 6 and 7 is large, i.e., they have a high first-order proximity, they should be represented closely to each other in the embedded space. At the same time, even though there is no direct link between vertices 5 and 6, they share many common neighbors, i.e., they have a high second-order proximity, and therefore should also be represented closely to each other. We expect that the consideration of the second order proximity effectively complements the sparsity of the first-order proximity and better preserves the global structure of the network.

\begin{wrapfigure}{r}{0.5\textwidth}
  \vspace{-38pt}
  \begin{center}
    \includegraphics[scale=.10]{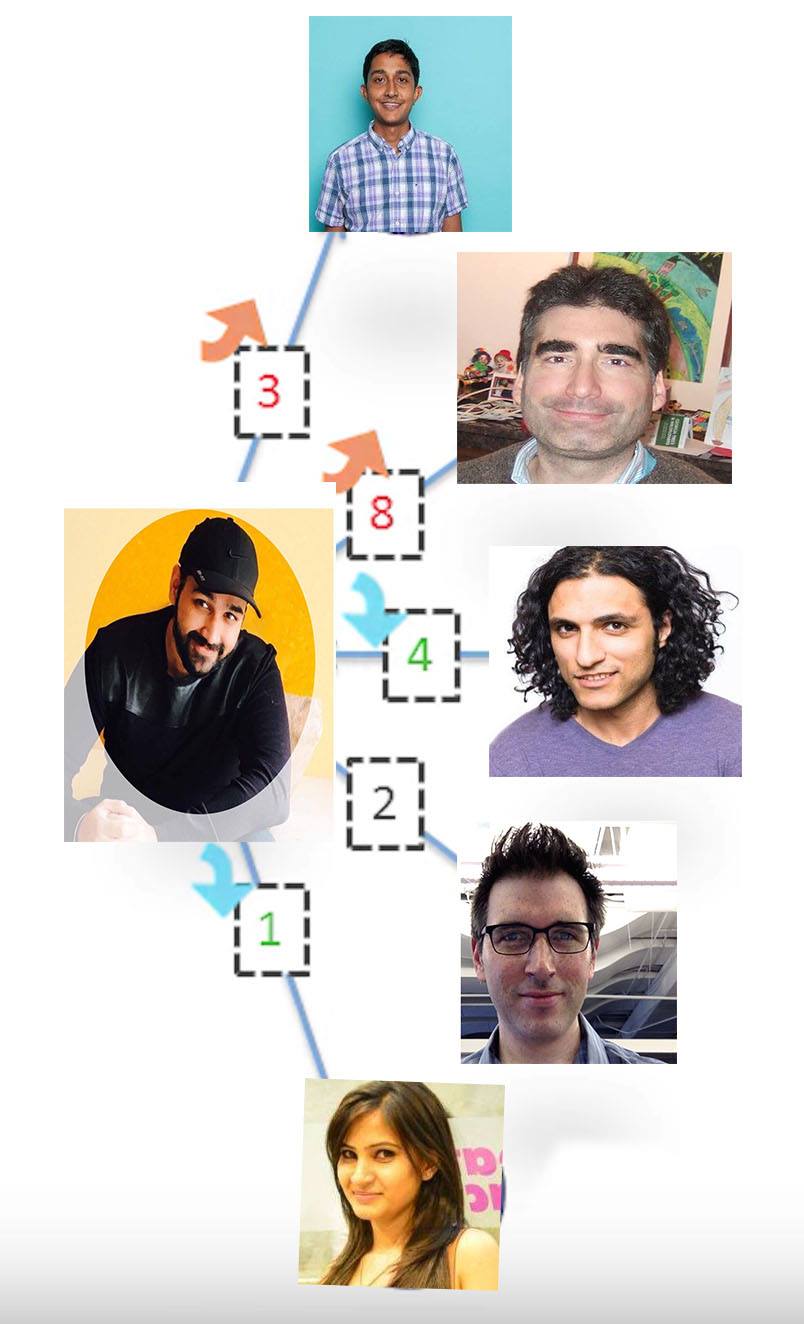}
  \end{center}
  \vspace{-10pt}
  \caption{Weighted Social Graph based on Interactions.}
  \vspace{-10pt}
\end{wrapfigure}
One of the first efforts in this direction was \textbf{LINE}\cite{tang:line}, a network embedding model, suitable for arbitrary types of information networks and efficiently scales to millions of nodes. The objective function is designed to preserve both the first-order and second-order proximities. Another approach is \textbf{Node2Vec}  \cite{grover:node2vec} which provides a flexible notion of  neighborhood and employs a biased random walk to efficiently explore diverse neighborhoods.
\subsubsection{Interaction-based Social Graph}
User and location based models can suffer from data sparsity, incase less data is available for user and also suffers from cold start problem. To overcome the deficiencies of the user-based model described in the previous section, we propose a joint model \textbf{soc2seq} whose objective combines \textbf{social embedding} with \textbf{conversation model}. The proposed model works online without retraining the entire model, since new user embeddings can be inferred using interaction logs and is robust to data sparsity as not many replies and posts are required per user. First, we discuss the social graph and how it is constructed at Yik Yak, and then we describe how the social graph can be used to learn low-dimensional embeddings and also explain how it can be combined with the prior models.
Once we have general location-based social graph, we used Node2Vec (which outperformed in link prediction task) to find to learn low-dimensional user embeddings. An important aspect to note is that in order to construct this model we did not make use of the textual content but rather we based it solely on user-to-user interactions.
\subsubsection{Building interaction graphs}
As mentioned above, one of the challenges we face at Yik Yak is the lack of an explicit notion of users being friends or following one another. Therefore, to build the social graph, we applied a bottom-up technique to leverage user interactions within the platform. Specifically, various interactions need to be combined and weighed to calculate the users affinity. The signals to be aggregated can be summarized in following categories:
\begin{itemize}[noitemsep,topsep=0pt]
  \item	\textbf{Profile view} (directed and binary): This graph is based on whether a user has viewed another user's profile.
\item	\textbf{Chat request} (directed and binary): This graph is based on whether a user has sent chat request to other user.
\item	\textbf{Comment}  (directed and weighted): The edges correspond to replies, where a user commented on another user's post. The weight is determined by the number of such interactions.
\item	\textbf{Like} (directed): The edges correspond to upvotes (likes), where a user liked another user's post. The weight is determined by the number of such interactions.
\item	\textbf{View} (directed-weak signal and weighted): The edges correspond to post views, where a user viewed another user's yak or comment. The weight is determined by the number of such interactions.
\end{itemize}
In practice, different graphs are used for different applications. For example, for a user recommendation or \enquote{people you may know} feature, the objective is to maximize chat requests and profile views. On the other hand, for the construction of personalized feed, the objective function would be a weighted combination of all five graphs. Node2Vec optimizes the following objective function, which maximizes the log-probability of observing a network neighborhood $N_S(u)$ for a node $u$ conditioned on its feature representation as given by $f$,
\begin{equation}
 	\begin{split}
 	& max_f  \sum_{u \in V }\log p(N_S(u)|f(u))
 	\end{split}
 \end{equation}
The main challenge here is the definition of the neighbors. The neighborhoods $N_S(u)$ are not  restricted to the immediate neighbors and can have vastly different structures depending on the sampling strategy $S$. Hence, the overall joint loss function of our model is given by:
\begin{equation}
 	\begin{split}
 	& loss_{total} = loss_{conversation}+loss_{social}\\
 	& = -\sum_{i=1}^{N}\ln p_{target} + SGD(node2vec\_walk)
 	\end{split}
\end{equation}\\
where SGD is the stochastic gradient descent on the Node2Vec random walk.

Now, due to the training complexity of running a random walk on the entire graph for every conversation, we first trained Node2Vec from the interaction graph and then used those embeddings in the user-based conversation models, as depicted in Fig. 2.  Moreover, Node2Vec parameters  such as \textbf{p} and \textbf{q} which are used to interplay between DFS and BFS are set to $1$, which were found to be the optimal values for downstream tasks such as chat link prediction.

Specifically for the reply suggestion task, we  used a concatenation of the \textbf{comment} and \textbf{like} graphs, so the embedding for user \textbf{Alice} is given by:
\begin{equation}
	\vec{user}_{\,Alice}= [\vec{comment}_{\,Alice} , \vec{like}_{\,Alice}] 
\end{equation}

\section{Training and Implementation}
Our work follows the footsteps of \cite{li:persona} work on persona-based conversation model.
\subsection{Training Protocols}
The overall training procedure, used across all of the methods, is as follows:
\begin{itemize}[noitemsep,topsep=0pt]
  \item 4 layer LSTM model with 1,000 hidden cells for each layer. 
  \item Batch size is set to 128.
  \item Learning rate is set to 1.0 with decay. 
  \item Parameters are initialized by sampling from the uniform distribution [$-0.1, 0.1$]. 
  \item Gradients are clipped to avoid gradient explosion with a threshold of 5. 
  \item Vocabulary size is limited to 100,000. 
  \item Dropout rate is set to 0.25.
\end{itemize}
\subsection{Decoding}
For the decoding phase, the N-best lists are generated using
the decoder with beam size $B = 200$. We set a
maximum length of $20$ for the generated candidates.
Decoding is performed  as follows: At each time step,
we first examine all $B\times B$ possible next-word candidates, and add all hypothesis ending with an EOS token to the N-best list. We then preserve the top-$B$
unfinished hypotheses and move to the next word.

\subsection{Dataset}
The dataset used for the training consists of yak (post) and comment (reply) pairs. We preprocessed the pairs such that each post contains at least 5 words and no explicit language. After preprocessing, we obtained about 10 million pairs which were randomly split in three parts, training, validation (development) and test (holdout, unseen) sets. This dataset encompassed 10,000 locations (at the county or city level) spread across 13 countries with a total of 100,000 unique users.

\subsection{Implementation}
\subsubsection{Training}
The source and target LSTMs use different sets of parameters.
We ran 20 epochs, and training took
roughly a week to finish on a g2.8xlarge AWS instance with 32 high frequency Intel Xeon E5-2670 (Sandy Bridge) processors
as well as 4 high-performance NVIDIA GPUs, each with 1,536 CUDA cores and 4 GB of video memory.

\subsubsection{Inference}
For inference, we used \textbf{Kubernetes}, an open-source system for automating deployment, scaling, and management of containerized applications. We used  Kubernetes with \textbf{TensorFlow Serving}, a high-performance, open-source serving system for machine learning models, to meet the computational intensity and scaling demands of our application. The server executes the TensorFlow graph to process every text suggestion request it receives. Kubernetes distributes the servicing of inference requests across a cluster using its External Load Balancer. Each pod in the cluster contains a TensorFlow Serving Docker image with the TensorFlow Serving Rest server and a trained SEQ2SEQ model. The model is represented as a set of files describing the shape of the TensorFlow graph, model weights, assets, and so on. Since everything is packaged together, we can dynamically scale the number of replicated pods using the Kubernetes Replication Controller to keep up with the service demands.
\section{Results and Discussion}
We have evaluated our models on Perplexity which guides probability of generating probable sequence as well as on Rouge score.
\subsection{Perplexity}
The typical measure used for comparing different models is perplexity, defined as
\begin{equation}
 	\begin{split}
 	& e^{-\sum_{i=1}^{N}\ln p_{target}} = e^{loss}
 	\end{split}
\end{equation}


\begin{table}
\centering
\caption{Perplexity of location- and user-based models.}\label{tab:results}
 \begin{tabular}{|c|c|c||} 
 \hline
  \textbf{Model} &  \textbf{Perplex.}     \\ [0.1ex] 
 \hline
LSTM (standard) & 79.1 \\ 
 \hline
LSTM (attention) & 77.2 \\
\hline
Location-based model (decoder) & 73.3  \\
\hline
Location-based model (decoder and encoder) & \textbf{72.7}\\
\hline
User-based model (decoder) & 79.6  \\
\hline
User-based model (decoder and encoder) & 80.7 \\
 \hline
Social user model (standard) & 72.4  \\
\hline
Social user model (tuned) & \textbf{70.9} \\
 \hline
\end{tabular}
 \end{table}
In Table \ref{tab:results}, we have summarized the results for our models  on test dataset and compared them with other techniques.
We observe that the location-based model results in a significant improvement (about $8\% $ reduction in perplexity), whereas the user-based model is outperformed by the baseline (LSTM) models (about $2\% $ increase in perplexity ). The data sparsity for users may explain this observation, as there is less data per user compared with per location. Moreover, the number of embeddings to be learned for the user-based model (1,00,000) is 10 times more than the corresponding number for location-based model (10,000). We further observe that the decoder-and-encoder model performs worse than the decoder model, which shows that the speaker information does not lead to better suggestions (i.e., replies) in an anonymous environment.

Models discussed earlier can suffer from data sparsity, in case less data is available for user and also suffers from cold start problem. This validates our hypothesis and our  motivation to build a novel model that incorporates user information in a more robust way.

As we can see in the results, using pretrained embeddings from the like and comment views of the social graph boosts the results, even without pretraining the user embeddings. The reason for such a boost is due to the fact that using the interaction graph we can find clusters of users in the embedding space that reply in a similar fashion. Furthermore, a significant gain is observed when we fine tune the user embeddings by using both social and conversation information. \\
\subsection{ROUGE}
\begin{table}
\centering
 \caption{ROUGE SCORE using our soc2seq model.}\label{tab:social-results}
 \begin{tabular}{|c|c|c|c|c|} 
 \hline
 \textbf{Model} &  \textbf{R-1} &  \textbf{R-2}   &  \textbf{R-L} \\ [0.01ex] 
 \hline
LSTM (standard) & 32.13& 11.73 & 29.26 \\ 
 \hline
LSTM (attention) & 34.17 &11.97 & 30.13 \\
\hline
Location-based model (decoder) & 35.67 & 12.41 & 31.13 \\
\hline
Location-based model (decoder and encoder) & 35.42 & 12.27 & 30.97 \\
\hline
User-based model (decoder and encoder) & 32.67 & 11.79 & 29.39 \\
\hline
Social user model (standard) & 36.8 & 12.83 & 31.24\\
\hline
Social user model (tuned) & \textbf{37.21} & \textbf{13.48} & \textbf{32.60} \\
 \hline
\end{tabular}
 \end{table}
Another evaluation metrics used is ROUGE\footnote{http://www.berouge.com/Pages/default.aspx} score. Since ROUGE score evaluates the suggestions based on n-gram match it looks into the style of writing as well as personalization of replies. We also used ROUGE score  as our evaluation
metric with standard options. The basic idea
of ROUGE is to count the number of overlapping
units between generated summaries and the reference
summaries, such as overlapped n-grams,
word sequences, and word pairs. 
 
F1 ROUGE scores are consistent with our earlier results with about 15.8\% improvement in ROUGE score for our social user model, which suggests that our model is able to understand  the idiosyncrasies of their conversational style as well as personalize the suggestions.
\section{Conclusion and future work}
We have presented a novel approach that jointly models the conversational and social aspects of user interactions. This model allows for intelligent agents that learn from both the content as well as the structure of user interactions and thus better emulate personalized behavior. It achieves a substantial gain in perplexity and ROUGE score for location-based  as well as social-based models. We have demonstrated that by encoding personas in distributed representations of conversation and social graphs, one can capture personal characteristics such as speaking style and background information.

This model represents a building block for future work, which includes making it more robust to unknown words by incorporating morphemes \cite{parry:morpho} or even character-level embeddings.
Also we plan to explore combining social graph with Li et al. 2016 approach \cite{li:rl} of using reinforcement learning as delayed rewards and policy gradient in order to incorporate diversity, ease of answering, and enforce semantic coherence. 
Yet another area might be to make sequential networks such as LSTM more intelligent as in HyperNetworks~\cite{ha:hn}, where a smaller network helps the main network to make intelligent decisions.

\section{Acknowledgments}
We thank Vijay Viswanathan, Yi Yang, Jacob Eisenstein, Hyokun Yun, Aaron Colak and Tomasz Jurczyk for their helpful discussion of this work.

\bibliographystyle{abbrvnat}
\bibliography{main}  







\appendix
\section{Practical outputs}
Having measured the performance of the system from a perplexity standpoint, it is also important to observe how it performs in practice. 

\subsection{Location-Based Examples}
We select five locations randomly for various posts to obtain the replies. Fig. 5 gives an example for the question \textit{Anyone wanna watch Netflix} and shows that different answers can be generated based on the location. From the replies it can be observed that \textit{Daredeveil} is popular among New Yorkers while \textit{Games of Thrones} is popular in London. Such flexibility could not have been possible in earlier models.
 \begin{figure}[!h]
\includegraphics[scale=.35]{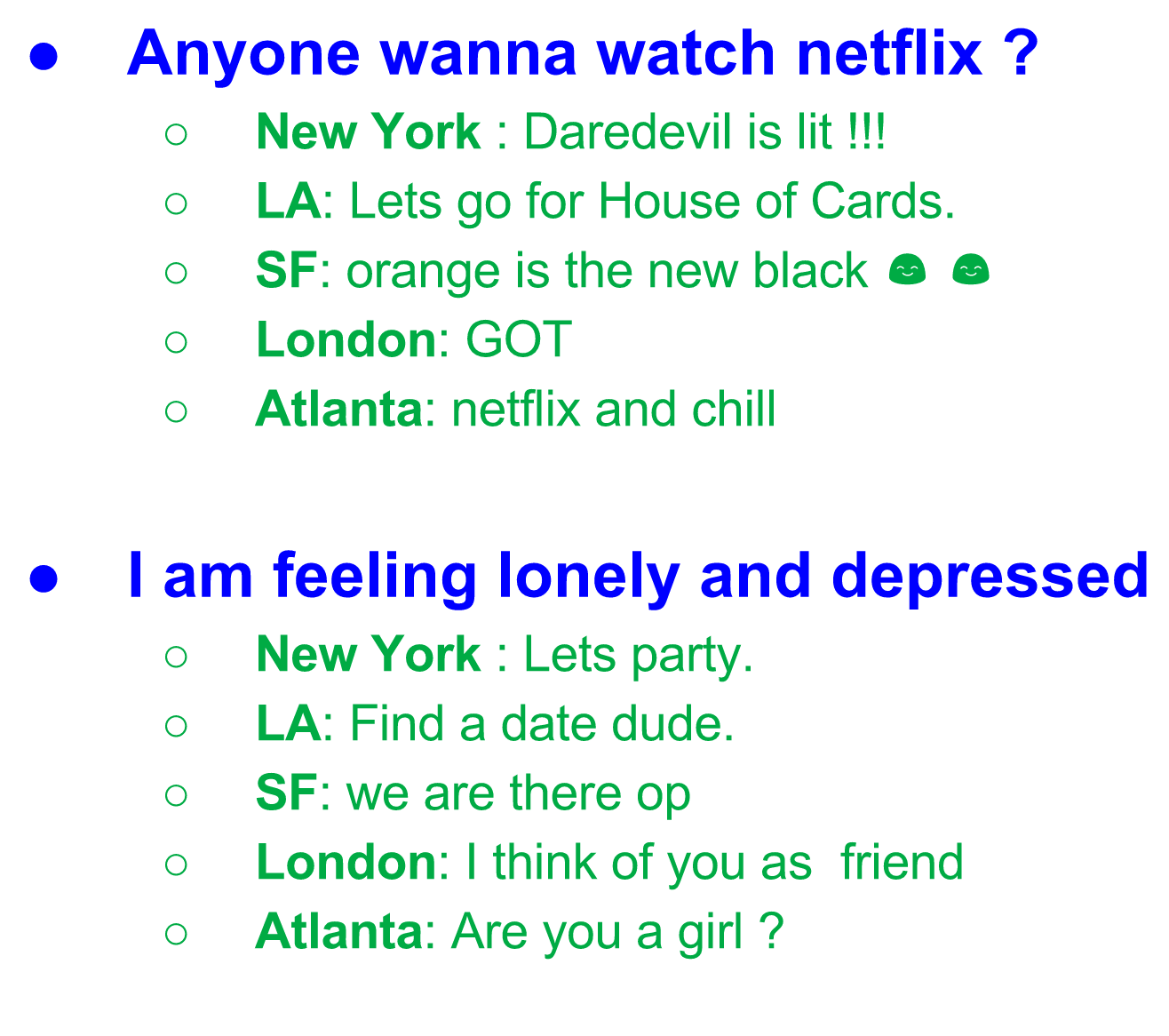}
\caption{Sample output of the location-based conversation model.}
\end{figure}
\subsection{Social-Graph-Based Examples}
\begin{figure}[!h]
\includegraphics[scale=.35]{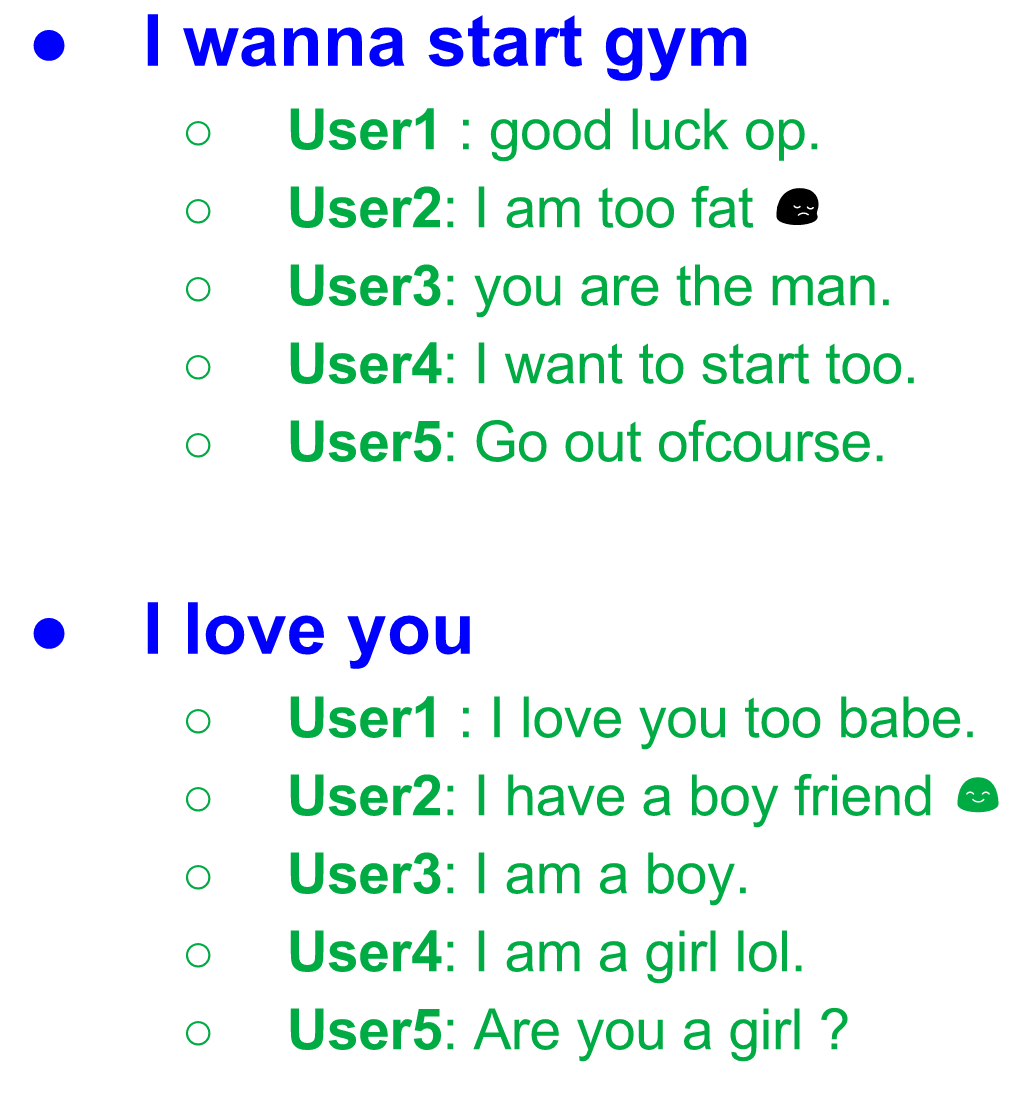}
\caption{Sample output of the soc2seq social conversation model.}

\end{figure}
For users, we randomly selected five people from the 100,000 users and evaluated their response on different sets of posts. As observed in Fig. 6, each of the have a unique but consistent personality, as observed in the way they reply. For example, User1 appears to be extrovert, and User4 is likely a female.\newline
\end{document}